# Novel Asymmetrical High-Resolution and High-Sensitivity Brain Dedicated PET system: Design optimization and performance evaluation

**Yuemeng Feng[1,+], Lisa Bläckberg[1,++] , Arkadiusz Sitek[1], and Hamid Sabet[1,*]**

[1] Department of Radiology, Massachusetts General Hospital & Harvard Medical School, Charlestown, 02129 Massachusetts, USA
[+] Present address: Universität zu Lübeck, DE
[++] Present address: Globus Medical, Methuen, 01844 Massachusetts, USA
*Author to whom any correspondence should be addressed

E-mail: yuemeng.feng@uni-luebeck.de, lisa.blackberg@gmail.com, asitek@mgh.harvard.edu, hsabet@mgh.harvard.edu

## Abstract

*Objective.* This study investigates the best achievable performance of a brain-dedicated PET system with high resolution and sensitivity by evaluating different detector configurations, while maintaining a practical system design suitable for dynamic brain I        imaging. *Approach.* Monte Carlo simulations were performed to evaluate system sensitivity and image quality under various timing resolutions (200 ps, 100 ps and 50 ps). The PET scanner geometry was optimized for human head imaging, featuring an elliptical cylindrical configuration with a neck cut-out, and front/back panels to enhance sensitivity and line of response (LOR) sampling. Detector configurations using LYSO:Ce crystals of varying thicknesses (15 mm and 20 mm) and depth of interaction (DOI) levels were simulated. Sensitivity was calculated using a point-like 511 keV back-to-back gamma source simulated at multiple locations within the field of view (FOV). Image reconstruction was conducted using list mode Maximum Likelihood Expectation Maximization (MLEM), assessing both a Derenzo-type phantom and a voxelized digital brain phantom. *Main results.* A location-dependent sensitivity ranging from 35.04% to 13.59% was achieved using a 20 mm thick LYSO:Ce crystal. Spatial resolution ranged from 0.8 mm to 1.5 mm within the FOV without applying resolution recovery techniques, measured using the FWHM of reconstructed hot rods. Previous results using 15 mm crystals with three DOI levels showed sensitivity between 23.42% and 15.99%, confirming the performance benefits of increased detector thickness and DOI capability. *Significance.* This study demonstrates the potential of a practical, brain-optimized PET system to achieve superior resolution and sensitivity for brain imaging. The findings offer valuable insights into optimal detector configurations, supporting the development of next-generation high-performance brain PET technologies.

## 1. Introduction

A brain-dedicated PET system with ultra-high resolution and sensitivity is evaluated in this study using Monte Carlo simulation toolkit Gate v9.3 [1]. The limitations of whole-body PET imaging systems in capturing small lesions and structures within the human brain, coupled with the excessive material costs associated with imaging larger fields of view (FOV) compared to a smaller FOV for brain imaging, makes it reasonable to develop brain-dedicated PET systems. With the objective of achieving high spatial resolution or increasing sensitivity of detection, numerous studies have been dedicated to advancing detector technologies, integrating CT/MRI information or improving attenuation correction and thus the image reconstruction [2-8]. Among the many brain-dedicated PET prototypes, significant efforts have been made for system design [5, 9, 10] and image reconstruction techniques [11-13], including attenuation correction [14-18], motion correction [19-22], resolution recovery [23-25] and AI-based methods [12, 16, 17, 23, 26]. In addition to PET alone imaging, several imaging devices have been developed that are capable of simultaneous multimodal acquisitions. The information provided by multimodal devices can be more comprehensive compared to a PET alone system. For example, PET/CT which combines anatomical structures with functional images [27], has shown greater accuracy than PET only systems [28]; PET/MRI combines functional and morphological images with good spatial resolution and image contrast of the soft tissue [29-31], and studies have explored the attenuation correction based on MRI information [15, 32-34]; the combination of PET, MRI and EGG (Electroencephalogram) could provide complementary anatomical, physiological, metabolic and functional information about the brain [35]. Although the combination with multiple imaging modalities is valuable, in this study, we focus on investigating a brain-dedicated PET system. Note that we use PET only systems existing in the literature as our reference.

A wearable brain PET system named PET-Hat is proposed in [36]. This system has a cylindrical shape with symmetric geometry. The transaxial resolution and axial resolution at the center of the FOV are approximately 4.0 mm FWHM and 3.5 mm FWHM, respectively. A helmet-shaped brain PET system is proposed in [37] and has shown a 1.97 mm FWHM of resolution in the FOV center. The NeuroEXPLORER human brain imager is reported to achieve a 1.64 mm FWHM radial spatial resolution, with a sensitivity of 11.8% at the FOV center [38]. Currently, the best reported resolution among existing brain-dedicated PET systems is 1.0 mm, as demonstrated in [5], surpassing the capabilities of the High resolution Research Tomography (HRRT) system [3, 4] which is widely used in clinical settings. However, this system uses a cylindrical geometry without coverage of the neck region, which can contain important information in spine, carotid arteries and thyroid. In conventional geometry designs, the best sensitivity performance is reported in [3] as 4.3%, with their best achievable resolution around 3 mm. A study on unconventional geometry design has shown a significant improvement on sensitivity [39], however the resolution remains around 2 mm. The table below summarizes the sensitivity and image resolution of several PET systems.

Performance of PET systems

| PET systems | Sensitivity | Resolution |
|---|---|---|
| NeuroEXPLORER [38] | 11.8% | 1.64 mm |
| HRRT [3,4] | 4.3% | 3 mm |
| PET-Hat [36] | 0.7% | 4.0-3.5 mm |
| helmet brain PET in [37] | 3.08% | 1.97 mm |
| brain PET in [5] | 3.4% | 1.0 mm |
| TRIMAGE PET [35] | 7.61% | 1.9-2.25 mm |

We have proposed a system that consists of an elliptical cylinder covering with a neck cut-out, supplemented by front and back panels to improve sensitivity and line-of-response (LOR) sampling, aiming to offer performance sufficient for dynamic brain imaging [40]. The design concept aims to maximize solid angle coverage of the human head. Our goal is to achieve both high resolution and high sensitivity for dynamic brain imaging. Through simulations we have previously demonstrated higher sensitivity than any existing brain PET systems by using a 15 mm thick staggered LYSO:Ce crystal with 3 depth of interaction (DOI) levels [40]. This work presents improvement on system design, and resolution-sensitivity metrics with detector configurations with varying DOI levels using a 20 mm thick detector and various values of coincidence timing resolution (CTR). A Derenzo phantom and an XCAT brain phantom are simulated and reconstructed using list mode MLEM with various detector configurations. A point-like source is simulated in air for measuring the sensitivity of system. The goal is to determine the best achievable system performance with varying detector configurations while keeping the design practical. Hot to Background Variability (HBV) and Hot to Background Contrast (HBC) are used for evaluating the image quality of reconstructed Derenzo phantom, and contrast recovery (CR) is used for evaluating the image quality of the reconstructed brain volume.

## 2. Methods

### 2.1 Brain-dedicated system

The proposed scanner consists of 12 elliptical cylinders, one back panel, and one front panel. The cylindrical part includes 4 fully populated rings (4x34 modules) and 8 partially populated rings (8x25 modules) to accommodate the patient's neck. The back panel contains of 102 modules, and the front panel has 88 modules with a 5x10 cm² opening for Virtual Reality (VR) camera and motion tracker. The eye-region opening also benefits claustrophobic patients. Total number of detector modules is 526, each consisting of 25x25x20 mm³ LYSO:Ce array with 1.5 mm pixel pitch one-one coupled with photodetector array with same pixel pitch. We plan to utilize SiPM or MPPC arrays such as Hamamatsu's S14826-3796 arrays with 1.3x1.3 mm² active area and 1.5 mm pitch. The visualization of the simulated brain PET system from Monte Carlo simulation toolkit Gate is presented in figure 1.

## 2.2 Monte Carlo simulation

We have performed a series of Monte Carlo simulations using Gate in order to assess the performance characteristics of the proposed PET scanner. For measuring the sensitivity, we simulated a point-like source emitting back-to-back gamma photons at 511 keV in air at different locations within the FOV, as shown in figure 1. The attenuation is not simulated in this study. Each source location is 7.5 cm away from its closest neighbor. The sensitivity is determined by dividing the number of detected coincidences by the number of total emissions. The detected coincidences are selected using a timing window of 1 ns. We implemented 20% energy resolution for the LYSO:Ce crystals. We applied an energy window of 350 keV to 650 keV for event selection.

To assess image quality within the FOV, we simulated a Derenzo phantom placed in seven regions of the system: the temporal lobes (region A, B), the parietal lobes (region C, F), the frontal lobes (region D), the brainstem (G) and one region near the patient's face (E). These placements are illustrated in figure 1. The Derenzo phantom as shown in figure 2 (a) includes hot rods with diameters of 5 mm, 4 mm, 2.5 mm, 1.5 mm, 1.0 mm, and 0.8 mm and is positioned at various locations in air for 30 minutes. For each distinct setting (i.e., different DOI, timing resolution, or source position), to reduce the computing time, one simulation was performed and split by acquisition time into 100 subsets, with each job corresponding to 18 seconds of simulated acquisition, enabling parallel computation on multiple CPU cores/threads without affecting the simulation results. The total detected counts remain at a similar level across different DOI and timing resolution settings but vary across different regions. The number of coincidences used in the reconstruction for each region is as follows: A: 9.2 M, B: 11 M, C: 14 M, D: 6.2 M, E: 8.2 M, F: 5.1 M, G: 5.9 M.

The total activity of the hot rods in a 1 cm length phantom was simulated as 9.446 MBq, with an average activity of 53 kBq/cc in each hot rod. Additionally, we reconstruct a brain volume with central slices presented in figure 2 (b), a Gaussian smoothing kernel is added to the volume. We have applied a Gaussian smoothing kernel to the reconstructed images in the Results section. The same smoothing kernel is also applied to the ground truth to ensure consistency in the visual comparison. The brain volume is obtained from an XCAT phantom, with the voxelized output saved and used in Gate, , with a simulated activity distribution of 1 Bq/cc per gray level.

The maximum value of grey matter region is set to 1.0, the minimum value of grey matter is set to 0.5, and for the white matter it is all set to 0.5. The digital phantom contains continuously distributed values between 0 and 1, not just three discrete values (0, 0.5, and 1). The total activity simulated in the whole brain was 49 MBq. Given the clinical administered dose of [$^{18}$F]FDG for adult brain imaging, which typically ranges from 125 to 250 MBq [41]. The brain uptake is typically about 6% of the injected dose, but can account for up to 20% of whole-body metabolism in the fasting state [49]. A 6% uptake corresponds to approximately 7.5–15 MBq in the brain, whereas a 20% uptake corresponds to 25–50 MBq. With a 20% uptake in the fasting state, the 49 MBq activity in the brain phantom is within the expected range. The simulated acquisition time was 20 minutes. For both the brain phantom and the Derenzo phantom, no scattering or attenuation was included in the simulation. Both phantoms were placed in air and were simulated to contain only back-to-back 511 keV gamma photons.

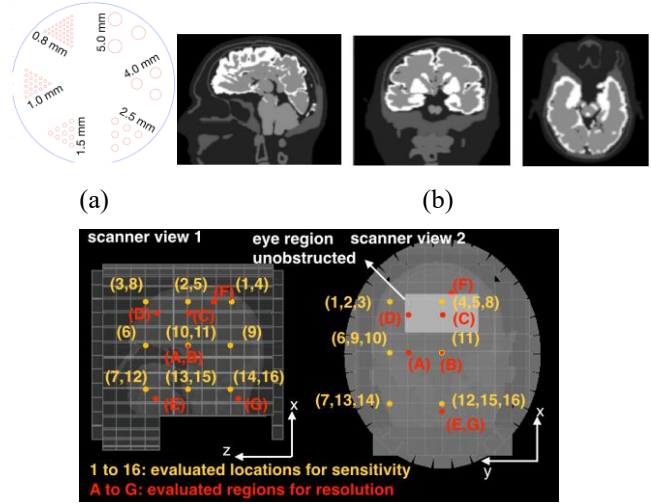

**Fig. 2:** (a) Central slices of the simulated Derenzo-type phantom, (b) the simulated 3D brain volume, (c) sensitivity simulations were performed at locations 1–16, Derenzo phantom reconstruction was performed at locations A–G.

Each image quality simulation (Derenzo phantom and brain volume) is repeated with three Coincidence Timing Resolution (CTR) values: 200 ps, 100 ps, 50 ps, and four DOI resolutions: 1 DOI (1x20 mm), 2 DOI (2x10 mm), 4 DOI (4x5 mm), 10 DOI (10x2 mm). The crystal size in each pixel is set to 1.4 mm [50] with 0.1 mm gap, and the pixel pitch is hence

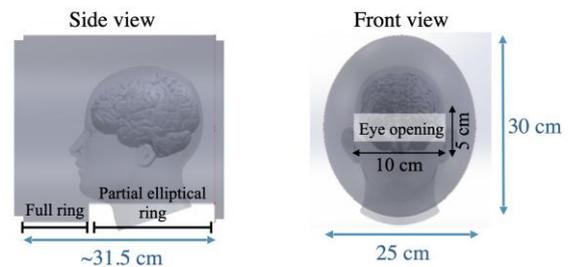

**Fig. 1:** Illustration of the brain-dedicated PET system: the eye region remains unobscured. The cylinder axis is aligned with the Z-axis, as shown in figure 1. The ellipse has radii of 125 mm and 150 mm. The cylinder length is 101.6 mm for the full ring and 203.2 mm for the partial coverage with a neck opening. The uncovered angular range of the partial ellipse is 120 degrees.

1.5 mm.

We have applied the following physics processes provided in Geant4 [42] during the simulations: PhotoElectric, Compton, RayleighScattering, ElectronIonisation, Bremsstrahlung, eMultipleScattering (e-), eMultipleScattering (e+), PositronAnnihilation. In the digitizer setting, scattered photons are kept, and we chose an energy resolution of 20% at 511 keV. The time resolution for singles was simulated as 35 ps, 71 ps or 141 ps. These correspond to TOF information of 50 ps, 100 ps [51], 200 ps [52]. A 1 ns time window was applied for calculating coincidences [51]. The random engine used is JamesRandom [43]. The crystal layer thickness varies from 2 mm to 20 mm, corresponding to DOI information ranging from 10 DOI to 1 DOI. Cross-scatter is not simulated in this study. Multiple interactions are not considered, either in the detector model or in the reconstruction. Regarding the DOI, detectors with different DOIs were simulated in GATE; therefore, the DOI binning is assumed to be perfect. DOI information is modeled deterministically in the simulation by subdividing each detector crystal into discrete DOI layers. For each interaction, the DOI layer is assigned based on the true interaction depth, and in the reconstruction the center position of the corresponding DOI layer is used. No additional DOI uncertainty or depth blur is applied.

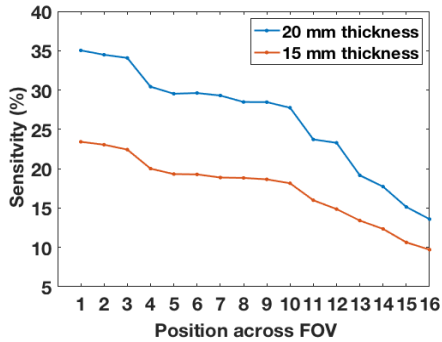

Fig. 3: Sensitivity across the FOV with two different crystal thicknesses: 15 mm, 20 mm.

### 2.3 Image reconstruction methods

List mode Maximum Likelihood Expectation Maximization (MLEM) is chosen for image reconstruction. Denoting the estimated dose distribution in a voxel $j$ within the reconstruction volume by $\lambda_j$, list mode MLEM aims to find $\lambda_j$ by calculating the sequence iteratively:

$$\lambda_j^{l+1} = \frac{\lambda_j^l}{s_j} \sum_{i=1}^{N} \frac{t_{ij}}{\sum_{k=1}^{M} t_{ik} \lambda_k^l},$$

where $l$ corresponds to the $l^{th}$ iteration. The iterative calculation is started by setting $\lambda_j^0$ to 1 for all voxels. The $t_{ij}$ corresponds to the possibility that two gamma photons generated from voxel $j$ are detected and recorded as coincidence $i$, which is one element of the system matrix. The $S_j$ corresponds to the possibility of the two-gamma emitted from voxel $j$ being detected, which is one element in the sensitivity matrix. The system matrix is calculated on-the-fly, and we are not using a pre-stored system matrix from measuring or simulating point sources. Denoting the detected locations of the two gamma photons as X and Y, the line of response (LOR) defined by X and Y as $L$, without adding the time of flight (TOF) information, in this study $t_{ij}$ is set to 1 if the distance between the center of the voxel $j$ and L is less than one voxel, and 0 if not. A Gaussian kernel is applied to $t_{ij}$ during each iteration to weight the LOR that crosses the voxel, with the standard deviation set to one voxel size. The TOF information is added to the calculation of $t_{ij}$ by applying a Gaussian kernel, and the $t_{ij}$ is calculated using the following equation: $t_{ij} = e^{-\frac{|V_j - V_t|^2}{2\sigma^2}}$, where $V_t$ is the calculated location on LOR using timing information, $V_j$ is the location of the voxel j on LOR, and $\sigma$ is the timing blurring width equaling to the CTR. In the reconstruction, TOF information is incorporated by weighting the system matrix on an event-by-event basis using a Gaussian kernel along the LOR. The kernel width is derived from the simulated CTR. Specifically, single-event timing resolutions of 35,71, 141 ps correspond to CTR values of 50, 100, 200 ps, assuming CTR=$\sqrt{2}\sigma$.

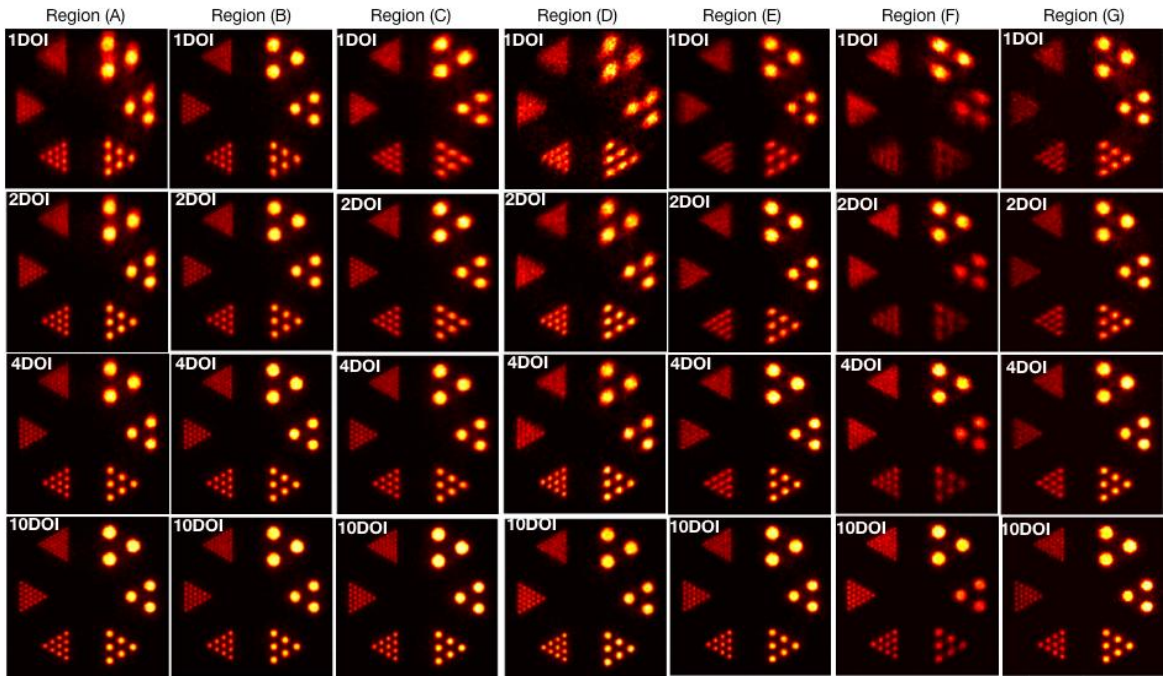

Fig. 4: Reconstruction results of the Derenzo phantom centered at region A to G at 40th iteration with different DOI information. The regions correspond to the temporal lobes (region A, B), the parietal lobes (region C, F), the frontal lobes (region D), the brainstem (G) and one region in the patient's face (E).

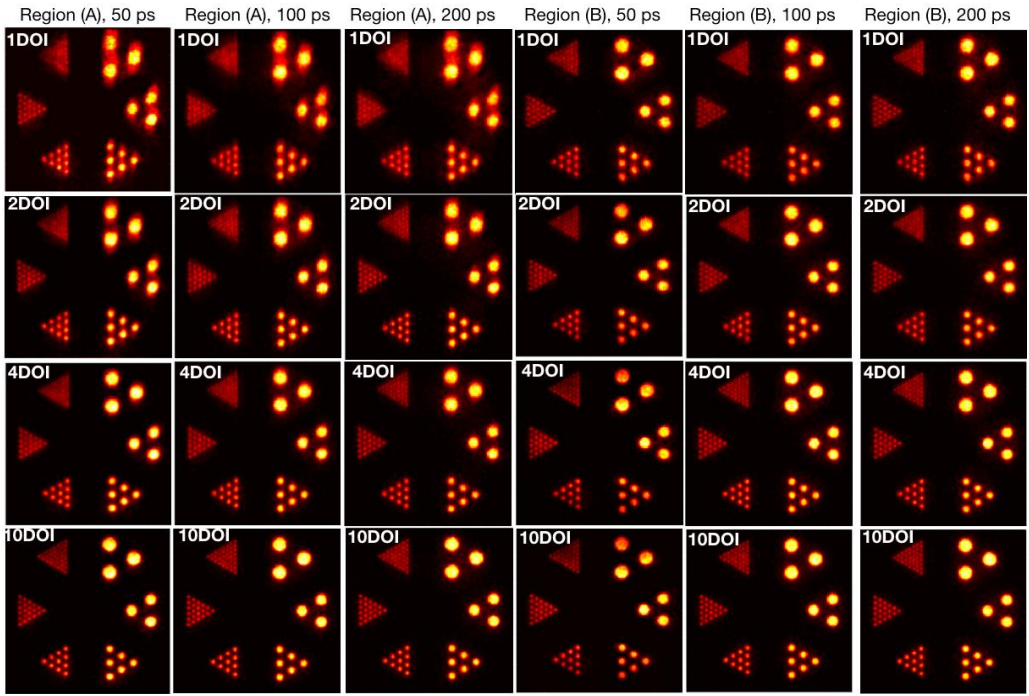

Fig. 5: Reconstruction of the Derenzo phantom centered at region A and B with varying TOF and DOI information.

All the reconstructed slices were obtained after 40 iterations of list-mode MLEM. The number of iterations was chosen because the reconstruction converged at the 40th iteration beyond which additional iterations introduce noise to the images. A mean square error (MSE) curve illustrating its variation with iteration number for one representative location has been included in the appendix. Forty iterations were selected to achieve a good visual trade-off between noise and contrast, and because the MSE tends to stabilize at this point, as shown in Fig. A.I. Although convergence may differ across positions within the FOV, we selected location F and its corresponding MSE to determine the number of iterations. This location is expected to exhibit the poorest image quality and the slowest convergence; therefore, the chosen number of iterations is sufficient to ensure stable convergence at other locations (e.g., location B). The objective of this study is to compare the impact of different system design parameters on reconstructed image quality. Consequently, a fixed number of 40 iterations was used for all configurations to limit the number of variables influencing the reconstruction and to ensure a fair comparison between system designs.

To evaluate image quality at different locations within the FOV, image quality metrics were compared at the same iteration number in order to avoid bias introduced by differing convergence behaviors. We used a voxel size of 0.4 mm³ for the reconstruction of the Derenzo phantom, and 1 mm³ for the reconstruction of the brain phantom. A post Gaussian smoothing is added to reconstructed results with standard deviation equal to one voxel.

## 2.4 Data analysis

To evaluate the image quality of the reconstructed Derenzo phantom, we used the Hot to Background Variability(HBV) and Hot to Background Contrast (HB) of the 1.0 mm rods as quality metrics. The value of HBV is calculated as: $HBV = \frac{\mu_{I_1} - \mu_{I_2}}{\sigma_{I_2}}$, where $\mu_{I_1}$ is the mean value of the region of the hot rods, $\mu_{I_2}$ the mean value of background, and $\sigma_{I_2}$ the standard deviation of the background. The HBC is calculated as the following equation: $HBC = \frac{\mu_{I_1} - \mu_{I_2}}{\mu_{I_2}}$.

For the evaluation of the reconstructed 3D brain volume, we applied the contrast recovery (CR) calculation as follows: $CR_{VOI}(\%) = 100 * \frac{\mu_{VOI} * \mu_{WMt}}{\mu_{WM} * \mu_{VOIt}}$, where $\mu_{VOI}$ corresponds to the mean value of the concentration of the reconstructed volume of interest (VOI), in this study VOI is the grey matter, and $\mu_{WM}$ corresponds to the mean value of the white matter, $\mu_{VOIt}, \mu_{WMt}$ correspond to ground truth. The phantom is simulated as a 3D image, where each voxel has a grey level. Voxels with a grey level greater than 0.5 are considered grey matter, while those with a grey level of 0.5 or less are considered white matter. The maximum grey level in the 3D image is 1.0. Given the simulation setting in this study, the $\frac{\mu_{VOIt}}{\mu_{WMt}}$ of ground truth is 1.594. Ideally, CR equals to 1.

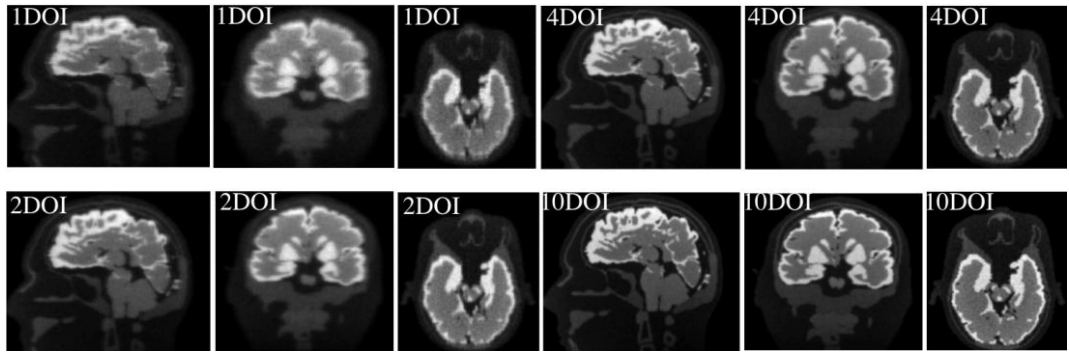

Fig. 6: Reconstruction of the brain phantom with 200 ps, using DOI information of 1DOI, 2DOI, 4DOI and 10DOI.

# 3. Results

## 3.1. sensitivity measurements

The sensitivity is calculated as coincidence counts divided by total number of emissions counts. The 16 evaluated locations are indicated in figure 1, covering the head and neck region. The sensitivity results comparing the simulated 15 mm-thick crystal simulated in previous study [40], with the 20 mm-thick crystal from this work are presented in table 1 in the appendix. The coincidences windows for event selection is 1 ns. For the 20 mm thickness crystal, the obtained sensitivity is 23.72% for center of geometric FOV (location 11), 28.48% for frontal lobe (location 8), 27.75% for cerebral cortex (location 10), and 13.59% for neck region (location 16). For the 15 mm thickness crystal, the obtained sensitivity is 15.99% for center of geometric FOV (location 11), 18.83% for frontal lobe (location 8), 18.15% for cerebral cortex (location 10), and 9.71% for neck region (location 16). The calculated sensitivity and the difference between the 20 mm thickness and 15 mm crystal thicknesses are shown in figure 3.

## 3.2. Reconstruction of Derenzo phantom

The reconstructed Derenzo phantom images are shown in figure 4 and figure 5. The line profiles of 1 mm rods located at the border of the 1 mm region are shown in Appendix.

## 3.3. Reconstruction of 3D brain volume

We show reconstruction of the simulated 3D brain volume in figure 6 and figure 7, with varying parameters. The reconstruction with 200 ps and varying DOI (1x20 mm, 2x10 mm, 4x5 mm, 10x2 mm) are shown in figure 6. The reconstruction of simulated data with 2 DOI and varying TOF information (50 ps, 100 ps, 200 ps) are shown in figure 7.

## 3.4. Image quality of the reconstruction

The image quality results for the reconstructed Derenzo phantom are presented in Tables 2 to 9 in the appendix, with evaluations performed using the 1 mm diameter rods. A comparison of image quality across different DOI and TOF levels is illustrated in figure 8.

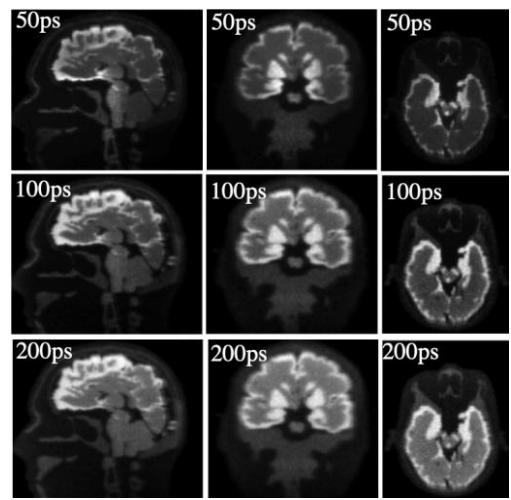

Fig. 7: Reconstruction of the brain phantom with 2 DOI information and TOF information 50 ps, 100 ps, 200 ps.

The contrast recovery (CR) of the reconstructed brain volume is summarized in table 10 in the appendix.

# 4. Discussion

In this study, we investigate the performance of a brain dedicated PET system with varying DOI information and TOF information. The idea is to evaluate the impacts of different parameters on the system performances and present the achievable system performance with the current geometry design and the current reconstruction package. The novelty of this study lies in the design of an asymmetric geometry specifically for human brain imaging. The results demonstrate good image resolution and relatively high sensitivity compared with existing brain PET systems [3,4,5,35,36,37,38]. The spatial resolution is evaluated only using 511 keV back-to-back events in air, without assessing the resolution in the presence of scattering medium. The effects of positron range and photon non-collinearity are also not considered, which leads to an optimistic estimation compared with realistic conditions.

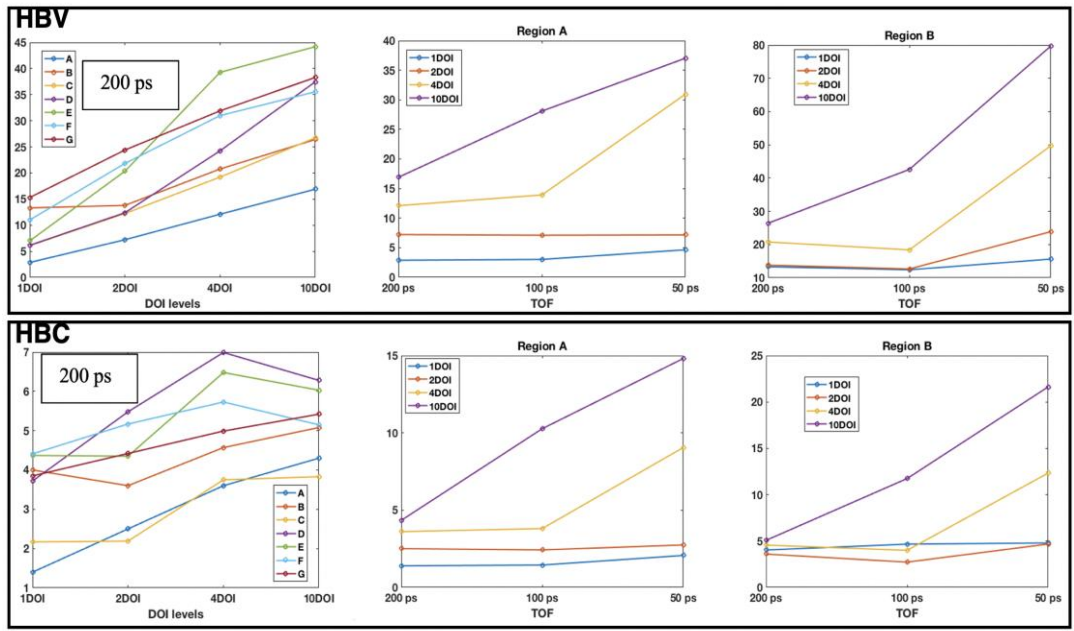

Fig. 8: HBV and HBC calculated from the reconstruction of the Derenzo phantom with different TOF and DOI levels. Top panels show HBC calculated from 200 ps reconstructions for phantom positions A-G, and from 200, 100, and 50 ps reconstructions with 1, 2, 4, and 10 DOI for phantom positions A and B. Bottom panels show the corresponding HBC results.

For evaluating the sensitivity within the FOV, we have simulated a point-like source at 16 different locations. Coincidences are calculated by using the timing information. The sensitivity is calculated by dividing the number of the coincidences by the emission counts of gamma photon. The evaluated locations together cover the brain and neck region. The crystal thickness is 20 mm. Results indicate that the current system could achieve higher sensitivity than existing systems [3,4,5,35,36,37,38]. For the locations tested, the sensitivity varies from 35.04% (location 1) to 13.59% (location 16). The closer the location is to the neck region, the smaller the difference between the two crystal thicknesses which is mainly due to limited solid angle.

With list-mode MLEM reconstruction applied to MC simulated data, a resolution of 1.0 mm can be achieved in the region A and B (temporal lobes region), and C (parietal lobes) for all the tested DOIs, without applying any resolution recovery method. The best achievable resolution in cerebral region (region F) at a radial distance of 10.6 cm from the center of the FOV is 1.0 mm with 4 DOI levels, and 0.8 mm resolution can be achieved with 10 DOI levels. In the same location, 1.5 mm rods can be resolved with 1 DOI. A resolution of 1.5 mm can be achieved in the regions of frontal lobes (region D) and brainstem (region G), with 1 DOI. The total activity simulated in Derenzo phantom is significantly lower than the clinical administered dose of [18F]FDG for adult brain imaging, which typically ranges from 125 to 250 MBq [41]. With a typical brain uptake of 6% in clinical imaging [49], the simulated activity in the Derenzo phantom falls within the expected range.

With the current system, we are able to obtain much higher sensitivity, allowing image reconstruction even with a limited simulated dose. With 10 DOI levels, 0.8 mm resolution can be achieved for all the tested locations.

For regions A and B, with 10 DOI levels, the image quality achieved with a 50 ps timing resolution is clearly superior to that with 100 ps or 200 ps (Figure 8), in terms of both HBV and HBC. However, with 1 DOI and 2 DOI levels, better TOF information yields little or no improvement, indicating, as expected, that the benefits of TOF are minimal when using a 20 mm thick crystal without DOI information. Performance improves with intermediate DOI resolution: with 4 DOI levels, TOF information provides noticeable gains, and at the 10 DOI levels, the benefit is the greatest. These results suggest that TOF should be prioritized when sufficient DOI resolution is available, as the two parameters work synergistically to improve image quality.

Results in figure 4 and HBVvalues in figure 8 shows that, 10 DOI levels improve the image quality. It is generally expected that increasing the number of DOI bins improves the spatial resolution of the reconstructed images. Increasing the DOI sampling lead to improvements in HBV across the tested regions. Based on the tests performed and the results obtained with the current system design, and without applying image resolution recovery methods, the combination of a CTR of 50 ps and 10 DOI layers provides the best performance.

For the simulated brain volume, we compared the effects of DOI information with TOF information simulated to 200 ps. Evident effects of the DOI information were shown from the calculated contrast recovery, as shown in table 10 in appendix. As expected, using 10 DOI yields the best results, closely approximating the ground truth. Using 4 or 10 DOI layers could be costly. For practical design considerations, 2 DOI layers are more realistic than 10, and since the target resolution is already achieved with 2 DOI, we used this configuration for the evaluation of the brain phantom. We tested the effects of the TOF information with DOI information set to 2 DOI, and the results indicate that better timing resolution improves contrast recovery. However, the reconstructed activity of white matter is decreased compared to the ground truth when using the 50 ps rather than 200 ps. This might be the reason why CR is improved when using better timing resolution.

The reconstructed results with the phantom simulated off-center indicate that the asymmetrical geometry affects image quality. It is well-known that MLEM reconstruction produces better image quality with more detected coincidences. However, in our simulated system, regions with higher sensitivity but located away from the center, such as location 3 in figure 1, which has the third highest sensitivity (34.08%) among the 16 tested locations, do not necessarily yield the best image quality. For example, with 1 DOI level, the reconstruction of the Derenzo phantom placed in region D (near location 3), shown in figure 4, does not exhibit the best image quality and appears slightly deformed. This observation highlights the need for high DOI resolution in asymmetrical geometries. In conventional cylindrical PET systems, improved DOI resolution enhances spatial resolution at the edge of the FOV, where the angular coverage and number of LORs are limited. Similarly, in asymmetrical geometries, non-uniform detector coverage leads to regions within the FOV with reduced and anisotropic LOR distributions.

Edge artifacts are observed in several line profiles (Appendix Figure A.2). The likely cause of the ringing artifacts, they appear more pronounced in regions with reduced angular coverage (e.g., the neck cut-out regions E and G) compared to central regions such as B, suggesting that angular undersampling is a contributing factor. Discrete DOI binning may also play a role, as regions sensitive to DOI information exhibit more artifacts when using 1 DOI layer compared to 10 DOI layers.

We will implement a more accurate calculation of the system matrix and incorporate the impacts of the asymmetric geometry into the LOR in future work. We chose the simple modeling of system matrix calculation as the reconstruction methods are not the focus of this manuscript. The block effects and positron range are not modeled in the current calculation. We noted that more accurate modeling of system matrix will further improve the image quality, such as methods using polar voxels and exact system matrix [44], using spatially variant resolution modeling [45]. Applying the point spread function (PSF) for image resolution recovery [46-48] will be considered for future work.

## 5. Conclusion

Seven different regions within the FOV are evaluated in this study for assessing the image quality, and 16 locations are evaluated for sensitivity of system. A resolution of 1.0 mm can be achieved in the cerebral cortex region using detectors with 2 DOI-levels, each with 10 mm thickness, while a resolution of 1.5 mm can be achieved in the brainstem region. The lowest image quality among the tested regions is in the brainstem, yet the resolution is still as good as 1.5 mm. For the simulated 20 mm thick crystal, the neck region, which includes the carotid artery and has useful signal for blood input function measurements, shows the lowest sensitivity at 13.59%. The best achievable image resolution is 0.8 mm in the cerebral cortex region using 10 DOI levels, and the best sensitivity is 35.04% in the parietal lobe region. A more accurate measurement of the PSF will be undertaken to further improve image quality. In addition to image reconstruction efforts, we are exploring multiple options to achieve the required DOI level, including fast waveform digitizing ASICs with built-in DOI measurement using light reflection ratio.

## Acknowledgements

The authors thank Dr. Salar Sajedi for his contributions to this project.

## References

1.   Jan, S., et al., *GATE: a simulation toolkit for PET and SPECT.* Physics in Medicine & Biology, 2004. **49**(19): p. 4543.
2.   Yamaya, T., et al., *Transaxial system models for jPET-D4 image reconstruction.* Physics in Medicine & Biology, 2005. **50**(22): p. 5339.
3.   Wienhard, K., et al., *The ECAT HRRT: performance and first clinical application of the new high resolution research tomograph.* IEEE Transactions on Nuclear Science, 2002. **49**(1): p. 104-110.

**Table 1.** Sensitivity simulated with a point-like source at varying locations within the FOV for crystals with 20 mm, 15 mm thickness

| Location | 1 | 2 | 3 | 4 | 5 | 6 | 7 | 8 | 9 | 10 | 11 | 12 | 13 | 14 | 15 | 16 |
|---|---|---|---|---|---|---|---|---|---|---|---|---|---|---|---|---|
| 20 mm, Sens(%) | 35.04 | 34.81 | 34.08 | 30.42 | 29.53 | 29.62 | 29.30 | 28.48 | 28.46 | 27.75 | 23.72 | 23.29 | 19.16 | 17.72 | 15.14 | 13.59 |
| 15 mm, Sens(%) | 23.42 | 23.05 | 22.43 | 20.01 | 19.32 | 19.28 | 18.88 | 18.83 | 18.65 | 18.15 | 15.99 | 14.87 | 13.41 | 12.35 | 10.64 | 9.71 |

4.  Schmand, M., et al., *Performance results of a new DOI detector block for a high resolution PET-LSO research tomograph HRRT.* IEEE Transactions on Nuclear Science, 1998. **45**(6): p. 3000-3006.

5.  Gaudin, É., et al., *Performance simulation of an ultrahigh resolution brain PET scanner using 1.2-mm pixel detectors.* IEEE transactions on radiation and plasma medical sciences, 2018. **3**(3): p. 334-342.

6.  Grogg, K.S., et al., *National Electrical Manufacturers Association and clinical evaluation of a novel brain PET/CT scanner.* Journal of Nuclear Medicine, 2016. **57**(4): p. 646-652.

7.  Tashima, H., H. Ito, and T. Yamaya. *A proposed helmet-PET with a jaw detector enabling high-sensitivity brain imaging.* in *2013 IEEE Nuclear Science Symposium and Medical Imaging Conference (2013 NSS/MIC).* 2013. IEEE.

8.  Li, H., et al., *Performance characteristics of the NeuroEXPLORER, a next-generation human brain PET/CT imager.* Journal of Nuclear Medicine, 2024. **65**(8): p. 1320-1326.

9.  Allen, M.S., M. Scipioni, and C. Catana, *New horizons in brain PET instrumentation.* PET clinics, 2024. **19**(1): p. 25-36.

10. LaBella, A., et al., *High-resolution depth-encoding PET detector module with prismatoid light-guide array.* Journal of Nuclear Medicine, 2020. **61**(10): p. 1528-1533.

11. Toussaint, M., R. Lecomte, and J.-P. Dussault, *Improvement of spatial resolution with iterative PET reconstruction using ultrafast TOF.* IEEE transactions on radiation and plasma medical sciences, 2020. **5**(5): p. 729-737.

12. Song, T.-A., et al., *PET image super-resolution using generative adversarial networks.* Neural Networks, 2020. **125**: p. 83-91.

13. Xu, H., et al., *Resolution modeling in projection space using a factorized multi-block detector response function for PET image reconstruction.* Physics in Medicine & Biology, 2019. **64**(14): p. 145012.

14. Mizuta, T., et al., *Initial evaluation of a new maximum-likelihood attenuation correction factor-based attenuation correction for time-of-flight brain PET.* Annals of Nuclear Medicine, 2022. **36**(4): p. 420-426.

15. Krokos, G., et al., *A review of PET attenuation correction methods for PET-MR.* EJNMMI physics, 2023. **10**(1): p. 52.

16. Hashimoto, F., et al., *Deep learning-based attenuation correction for brain PET with various radiotracers.* Annals of Nuclear Medicine, 2021. **35**: p. 691-701.

17. Lee, J.S., *A review of deep-learning-based approaches for attenuation correction in positron emission tomography.* IEEE Transactions on Radiation and Plasma Medical Sciences, 2020. **5**(2): p. 160-184.

18. Shiri, I., et al., *Direct attenuation correction of brain PET images using only emission data via a deep convolutional encoder-decoder (Deep-DAC).* European radiology, 2019. **29**: p. 6867-6879.

19. Onishi, Y., et al., *Performance evaluation of dedicated brain PET scanner with motion correction system.* Annals of Nuclear Medicine, 2022. **36**(8): p. 746-755.

20. Spangler-Bickell, M.G., et al., *Evaluation of data-driven rigid motion correction in clinical brain PET imaging.* Journal of Nuclear Medicine, 2022. **63**(10): p. 1604-1610.

21. Lu, Y., et al., *Data-driven motion detection and event-by-event correction for brain PET: comparison with Vicra.* Journal of Nuclear Medicine, 2020. **61**(9): p. 1397-1403.

22. Zeng, T., et al., *Markerless head motion tracking and event-by-event correction in brain PET.* Physics in Medicine & Biology, 2023. **68**(24): p. 245019.

23. Liu, J., et al., *Artificial intelligence-based image enhancement in PET imaging: noise reduction and resolution enhancement.* PET clinics, 2021. **16**(4): p. 553.

24. Song, T.-A., et al., *Super-resolution PET imaging using convolutional neural networks.* IEEE transactions on computational imaging, 2020. **6**: p. 518-528.

25. Song, T.-A., et al., *PET image deblurring and super-resolution with an MR-based joint entropy prior.* IEEE transactions on computational imaging, 2019. **5**(4): p. 530-539.

26. Sanaat, A., et al., *Deep-TOF-PET: Deep learning-guided generation of time-of-flight from non-TOF brain PET images in the image and projection domains.* Human brain mapping, 2022. **43**(16): p. 5032-5043.

27. Townsend, D.W., et al., *PET/CT today and tomorrow.* Journal of Nuclear Medicine, 2004. **45**(1 suppl): p. 4S-14S.

28. Aide, N., et al. *Advances in PET/CT technology: an update.* in *Seminars in nuclear medicine.* 2022. Elsevier.

29. Delso, G. and S. Ziegler, *PET/MRI system design.* European journal of nuclear medicine and molecular imaging, 2009. **36**: p. 86-92.

30. Jadvar, H. and P.M. Colletti, *Competitive advantage of PET/MRI.* European journal of radiology, 2014. **83**(1): p. 84-94.

31. Ehman, E.C., et al., *PET/MRI: where might it replace PET/CT?* Journal of Magnetic Resonance Imaging, 2017. **46**(5): p. 1247-1262.

32. Yang, X., et al., *MRI-based attenuation correction for brain PET/MRI based on anatomic signature and machine learning.* Physics in Medicine & Biology, 2019. **64**(2): p. 025001.

33. Chen, Y. and H. An, *Attenuation correction of PET/MR imaging.* Magnetic Resonance Imaging Clinics, 2017. **25**(2): p. 245-255.

34. Izquierdo-Garcia, D., et al., *Comparison of MR-based attenuation correction and CT-based attenuation correction of whole-body PET/MR imaging.* European journal of nuclear medicine and molecular imaging, 2014. **41**: p. 1574-1584.

35. Masturzo, L., et al., *Monte carlo characterization of the trimage brain PET system.* Journal of Imaging, 2022. **8**(2): p. 21.

36. Yamamoto, S., et al., *Development of a brain PET system, PET-Hat: a wearable PET system for brain research.* IEEE Transactions on Nuclear Science, 2011. **58**(3): p. 668-673.

37. Wang, T., et al., *Design and simulation of a helmet brain PET system.* Nuclear Instruments and Methods in Physics Research Section A: Accelerators, Spectrometers, Detectors and Associated Equipment, 2020. **978**: p. 164470.

38. Li, H., et al., *Performance Characteristics of the NeuroEXPLORER, a Next-Generation Human Brain PET/CT Imager.* Journal of Nuclear Medicine, 2024.

39. Shi, H., et al., *Design study of dedicated brain PET with polyhedron geometry.* Technology and Health Care, 2015. **23**(s2): p. S615-S623.

40. Bläckberg, L., et al. *High sensitivity and high resolution dynamic brain-dedicated TOF-DOI PET scanner.* in *2020*

*IEEE Nuclear Science Symposium and Medical Imaging Conference (NSS/MIC)*. 2020. IEEE.

41. Guedj, E., et al., *EANM procedure guidelines for brain PET imaging using [18 F] FDG, version 3.* European journal of nuclear medicine and molecular imaging, 2022: p. 1-20.

42. Agostinelli, S., et al., *GEANT4—a simulation toolkit.* Nuclear instruments and methods in physics research section A: Accelerators, Spectrometers, Detectors and Associated Equipment, 2003. **506**(3): p. 250-303.

43. Sepehri, F., M. Hajivaliei, and H. Rajabi, *Selection of random number generators in GATE Monte Carlo toolkit.* Nuclear Instruments and Methods in Physics Research Section A: Accelerators, Spectrometers, Detectors and Associated Equipment, 2020. **973**: p. 164172.

44. Ansorge, R. *List mode 3D PET reconstruction using an exact system matrix and polar voxels*. in *2007 IEEE Nuclear Science Symposium Conference Record*. 2007. IEEE.

45. Bickell, M.G., L. Zhou, and J. Nuyts, *Spatially variant resolution modelling for iterative list-mode PET reconstruction.* IEEE transactions on medical imaging, 2016. **35**(7): p. 1707-1718.

46. Varrone, A., et al., *Advancement in PET quantification using 3D-OP-OSEM PSF reconstruction with the HRRT.* Neuroimage, 2008. **41**: p. T85.

47. Ahrari, S., et al., *Implementing the point spread function deconvolution for better molecular characterization of newly diagnosed gliomas: A dynamic 18F-FDOPA PET radiomics study.* Cancers, 2022. **14**(23): p. 5765.

48. Dutta, J., et al. *PET point spread function modeling and image deblurring using a PET/MRI joint entropy prior*. in *2015 IEEE 12th international symposium on biomedical imaging (ISBI)*. 2015. IEEE.

49. DShammas, A., Lim, R. and Charron, M., 2009. Pediatric FDG PET/CT: physiologic uptake, normal variants, and benign conditions. *Radiographics*, *29*(5), pp.1467-1486.

50. Jiang, W., Chalich, Y. and Deen, M.J., 2019. Sensors for positron emission tomography applications. *Sensors*, *19*(22), p.5019.

51. Schaart DR, Seifert S, Vinke R, van Dam HT, Dendooven P, Löhner H, Beekman FJ. LaBr3: Ce and SiPMs for time-of-flight PET: achieving 100 ps coincidence resolving time. Physics in Medicine & Biology. 2010 Mar 19;55(7):N179.

52. Borghi, G., Tabacchini, V., Bakker, R. and Schaart, D.R., 2018. Sub-3 mm, near-200 ps TOF/DOI-PET imaging with monolithic scintillator detectors in a 70 cm diameter tomographic setup. *Physics in Medicine & Biology*, *63*(15), p.155006.

**Appendix**

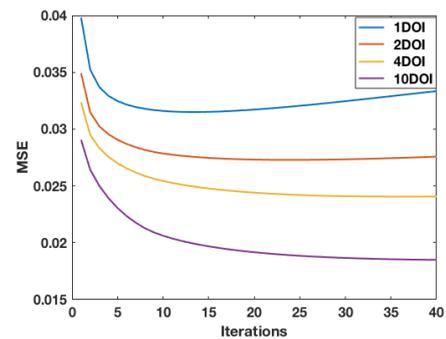

Fig. A.I: MSE vs iterations at location F.

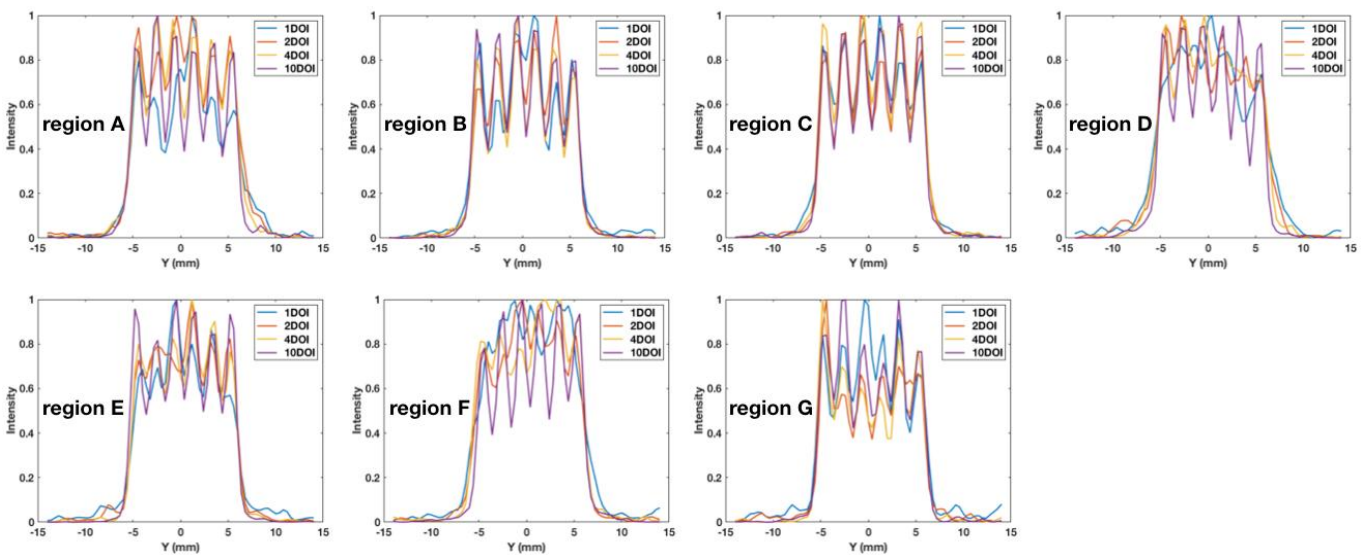

Fig. A.2: Line profile of 1 mm rods indicated in the reconstructed images shown in figure 4.

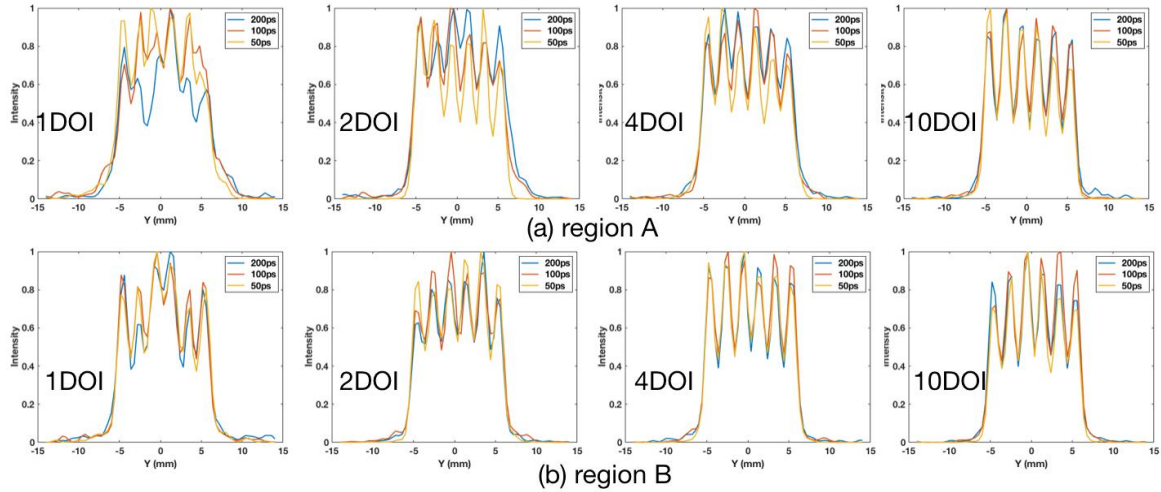

Fig. A.3: Line profile of 1 mm rods indicated in the reconstructed images shown in figure 5.

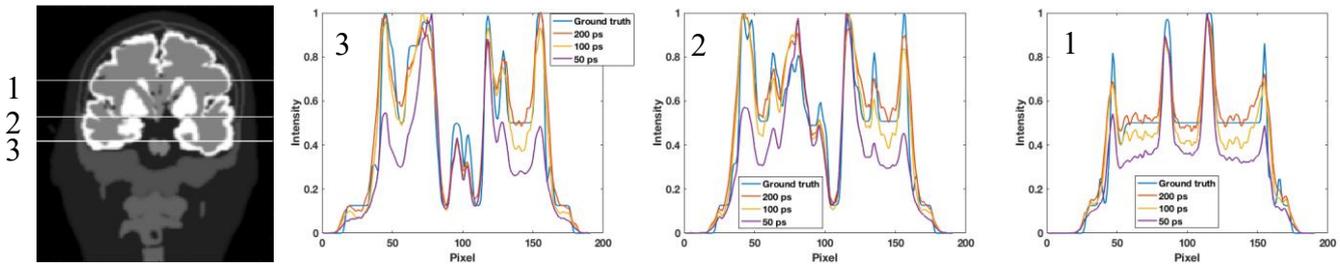

Fig. A.4: Line profiles of the reconstructed brain volume with 2DOI and different TOF information.

**Table 2.** HBV and HBC with 1 DOI, 200 ps.

| Region | A | B | C | D | E | F | G |
|--------|------|-------|------|------|------|-------|-------|
| HBV | 2.87 | 13.32 | 6.09 | 6.15 | 7.00 | 11.00 | 15.33 |
| HBC | 1.40 | 4.04 | 2.17 | 3.72 | 4.37 | 4.41 | 3.85 |

**Table 6.** HBV and HBC with 1 DOI, 100 ps, 50 ps.

| Regions, TOF (ps) | A, 100 | A, 50 | B, 100 | B, 50 |
|-------------------|--------|-------|--------|-------|
| HBV | 3.01 | 4.64 | 12.41 | 15.62 |
| HBC | 1.45 | 2.07 | 4.67 | 4.80 |

**Table 3.** HBV and HBC with 2 DOI, 200 ps.

| Regions | A | B | C | D | E | F | G |
|---------|------|-------|-------|-------|-------|-------|------|
| HBV | 7.22 | 13.80 | 12.24 | 12.39 | 20.34 | 21.83 | 24.4 |
| HBC | 2.51 | 3.60 | 2.19 | 5.48 | 4.35 | 5.17 | 4.42 |

**Table 7.** HBV and HBC with 2 DOI, 100 ps, 50 ps.

| Regions, TOF (ps) | A, 100 | A, 50 | B, 100 | B, 50 |
|-------------------|--------|-------|--------|-------|
| HBV | 7.10 | 7.16 | 12.66 | 23.87 |
| HBC | 2.43 | 2.75 | 2.73 | 4.68 |

**Table 4.** HBV and HBC with 4 DOI, 200 ps.

| Region | A | B | C | D | E | F | G |
|--------|-------|-------|-------|-------|-------|------|------|
| HBV | 12.11 | 20.77 | 19.22 | 24.25 | 39.28 | 31.0 | 31.9 |
| HBC | 3.61 | 4.57 | 3.75 | 6.99 | 6.48 | 5.73 | 4.99 |

**Table 8.** HBV and HBC with 4 DOI, 100 ps, 50 ps.

| Regions, TOF (ps) | A, 100 | A, 50 | B, 100 | B, 50 |
|-------------------|--------|-------|--------|-------|
| HBV | 13.93 | 30.95 | 18.40 | 49.71 |
| HBC | 3.81 | 9.05 | 4.00 | 12.33 |

**Table 5.** HBV and HBC with 10 DOI, 200 ps.

| Region | A | B | C | D | E | F | G |
|--------|------|-------|-------|-------|-------|-------|------|
| HBV | 16.9 | 26.41 | 26.74 | 37.47 | 44.19 | 35.55 | 38.3 |
| HBC | 4.34 | 5.08 | 3.83 | 6.28 | 6.03 | 5.15 | 5.42 |

**Table 9.** HBV and HBC with 10 DOI, 100 ps, 50 ps.

| Regions, TOF (ps) | A, 100 | A, 50 | B, 100 | B, 50 |
|-------------------|--------|-------|--------|-------|
| HBV | 28.09 | 37.04 | 42.60 | 79.70 |
| HBC | 10.27 | 14.81 | 11.74 | 21.62 |

**Table 10.** CR of the brain volume.

| DOI, TOF | 1 DOI, 200 ps | 2 DOI, 200 ps | 4 DOI, 200 ps | 10 DOI, 200 ps | 2 DOI, 100 ps | 2 DOI, 50 ps |
|---|---|---|---|---|---|---|
| CR(%) | 27.2 | 26.7 | 72.1 | 77.8 | 33.8 | 44.5 |

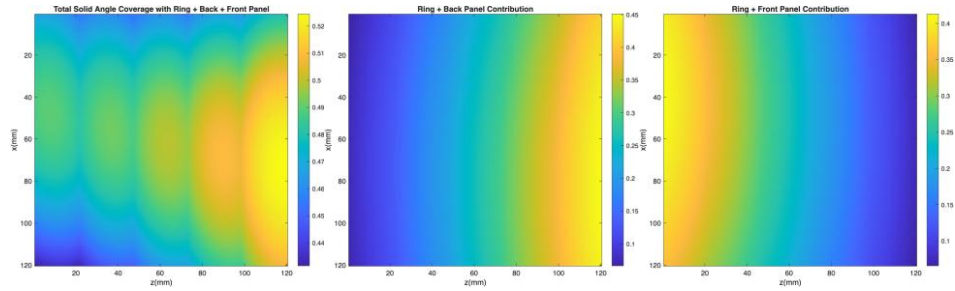

Fig. A.5: Analytical solid-angle coverage of the entire system and the individual contributions from the front and back panels on a centered 12 cm × 12 cm plane.